\documentclass[journal]{IEEEtran}
\usepackage{amsmath} 
\usepackage{amssymb}  
\usepackage{hyperref}
\usepackage{multirow}
\usepackage{subcaption}
\usepackage[pdftex]{graphicx}
\usepackage[]{algorithm2e}
\usepackage{lipsum}
\usepackage{enumerate}
\usepackage{tabularx}
\usepackage{color}
\usepackage[english]{babel}
\usepackage{blindtext}

\usepackage{float}

\usepackage{array}
\newcolumntype{L}[1]{>{\raggedright\let\newline\\\arraybackslash\hspace{0pt}}m{#1}}
\newcolumntype{C}[1]{>{\centering\let\newline\\\arraybackslash\hspace{0pt}}m{#1}}
\newcolumntype{R}[1]{>{\raggedleft\let\newline\\\arraybackslash\hspace{0pt}}m{#1}}

\ifCLASSINFOpdf
\else
\fi

\begin{document}
%
%
%
%


\title{Looking at Hands in Autonomous Vehicles:\\A ConvNet Approach using Part Affinity Fields}

\author{Kevan Yuen and Mohan M. Trivedi\\
  \emph{LISA: Laboratory for Intelligent \& Safe Automobiles}\\
  \emph{University of California San Diego}\\
  \emph{kcyuen@eng.ucsd.edu, mtrivedi@eng.ucsd.edu}\\
}

\maketitle

\begin{abstract}
In the context of autonomous driving, where humans may need to take over in the event where the computer may issue a takeover request, a key step towards driving safety is the monitoring of the hands to ensure the driver is ready for such a request. This work, focuses on the first step of this process, which is to locate the hands. Such a system must work in real-time and under varying harsh lighting conditions. This paper introduces a fast ConvNet approach, based on the work of original work of OpenPose by Cao, et. al. \cite{cao2017realtime} for full body joint estimation. The network is modified with fewer parameters and retrained using our own day-time naturalistic autonomous driving dataset to estimate joint and affinity heatmaps for driver \& passenger's wrist and elbows, for a total of 8 joint classes and part affinity fields between each wrist-elbow pair. The approach runs real-time on real-world data at 40 fps on multiple drivers and passengers. The system is extensively evaluated both quantitatively and qualitatively, showing at least 95\% detection performance on joint localization and arm-angle estimation.
\end{abstract}


%
\IEEEpeerreviewmaketitle

\section{Introduction}
Autonomous vehicles has been on a rise since the first release of commercially available luxury vehicle capable of autonomous driving in 2015 \cite{tesla2015release1}\cite{tesla2015release2} and the recent release of a more affordable autonomous vehicle in 2017\cite{teslamodel3release1}. The commercially available Tesla autopilot feature is considered to be between SAE levels 2-3 \cite{SAEJ3016} at the time of this writing. Level 2 requires that the driver ``supervises the driving automation system and intervenes as necessary to maintain safe operation of the vehicle'' and ``determines whether/when engagement \& disengagement of the driving automation system is appropriate''. With the vehicle in control of the driving, the driver is then responsible for two important tasks: 1) monitoring the system to ensure it is operating correctly, and 2) be ready to take over it is not operating correctly \cite{bainbridge1983ironies}.
\begin{figure}
  \centering
  \begin{tabular}{c}
        \includegraphics[width=0.47\textwidth]{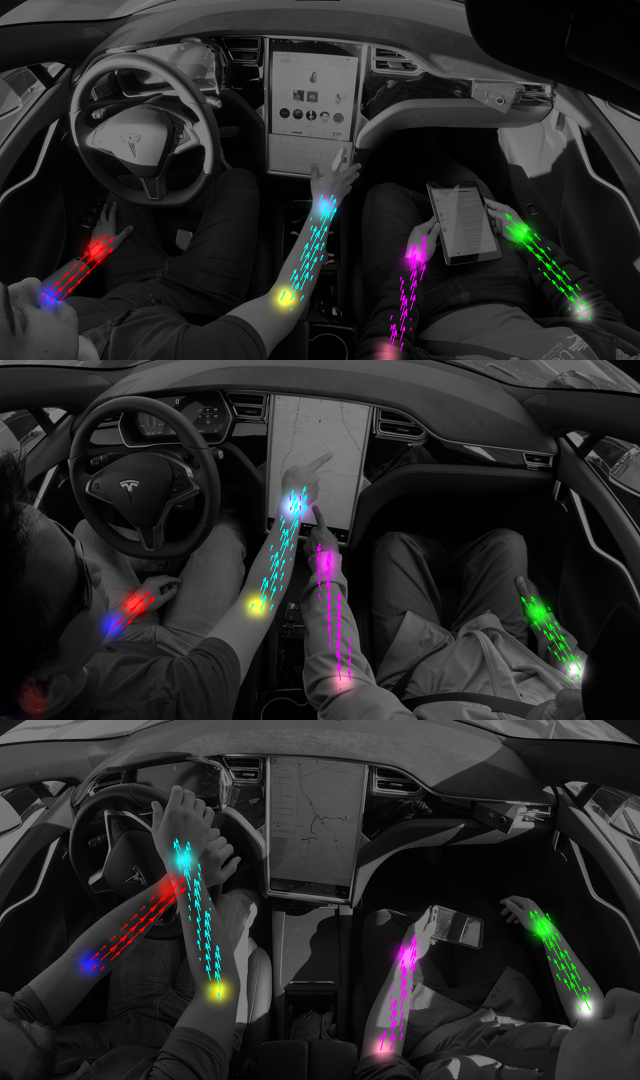}
  \end{tabular}
  \caption{Visualization of the network output showing the localization of driver \& passenger arms. The overlaid circular heatmaps correspond to the part location of the elbow and wrist for each arm, where each color is a different channel output corresponding to each of the driver and passenger arms. The arrows show the Part Affinity Fields of each arm, pointing from elbow to wrist.}
  \vspace{-5mm}
  \label{fig:intro_fig}
\end{figure}
Many vehicle manufacturers are beginning to catch up by releasing commercially available vehicles with autonomous capabilities. A recent vehicle manufacturer has recently claimed that they are producing the first commercially available vehicle with level 3 autonomy \cite{AudiA8TrafficJamPilot}. Level 3 no longer requires the driver to continuously supervise the automation of the vehicle. This means that the vehicle is responsible for determining when the driver should take over and requires the driver ``is receptive to a request to intervene and responds by performing the dynamic driving task in a timely manner''.

Driver distraction has been found to be a main contributor to the cause of vehicle accidents \cite{klauer2010analysis}, where drivers may be involved in secondary tasks such as: cell-phone usage, printed material reading, eating, drinking, grooming, or interacting with non-essential vehicle devices. Secondary tasks were found delay driver's take-over time \cite{eriksson2017takeover}\cite{mok2017tunneled} during critical situations when the driver must take back control of the vehicle when issued a take-over-request. In \cite{gold2016taking}, a study of surround traffic density and take-over performance showed that ``the presecense of traffic in takeover situations led to longer takeover times and worse take-over quality in the form of shorter time to collision and more collisions.'' With autonomous vehicles, driver monitoring becomes increasingly important as the driver may feel more comfortable to perform secondary tasks over time. With the system having knowledge of the driver state, it may notify a distracted driver with both sound and visual alerts to pay more attention to the road when there is a higher risk situation determined by the system using outside information, such as surround-traffic and road conditions, in order to improve a possible takeover performance. If the driver has already been paying attention, a smaller alert can be used to notify the driver instead.

In a survey \cite{poll2011most}, 37\% of drivers admitted to sending/receiving text messages and 86\% reported eating and drinking. The amount of time driver's eyes off-road were increased by up to 400\% and made 140\% more incorrect lane changes when texting \cite{world2011mobile}. Hands are an important cue to determining the driver's attention as many secondary activities such as eating, drinking, texting, and GPS system interaction require the use of hands. The research in this paper makes the following four contributions. (1) Adaptation of an existing ConvNet architecture with parameter reduction for accurate and efficient monitoring of vehicle occupant hand activites during autonomous driving. (2) Novel augmentation methods for reducing annotation time and increasing variability in training data to reduce overfitting. (3) Extensive quantitative and qualitative analysis and evaluation of the ConvNet approach developed in this work using a test vehicle capable of sophisticated autonomous driving modes for extensive trials in the real-world environment. (4) Releasing of annotated data for independent evaluation by other research teams (pending approval by the UCSD Technology Transfer Office).

This paper focuses on building a robust framework for determining the location of driver and passenger hands as a first but important step towards attention and distraction monitoring.

\section{Related Studies}

The research of hand detection in vehicle have been studied for about a decade. In this section we highlight a selected set of relevant studies. A study by Ohn-Bar \& Trivedi cite{ohn2013vehicle} detects the presence of hands in pre-defined regions using a Linear SVM comparing different image descriptors such as skin-based detectors, HOG \cite{dalal2005histograms}, and GIST \cite{oliva2001modeling}. Das et. al. \cite{das2015performance} explores the use of an ACF detector \cite{dollar2010fastest} for hands and also presented a challenging video-based dataset for driver hands with varying camera positions and angles. Our work is not evaluated on this dataset since this work focuses on a specific vehicle and camera position/angle shown in Fig. \ref{fig:intro_fig}, which provides optimal performance when trained and tested on the same vehicle and thus maximum safety. More recent work on hands have focused on Convolutional Neural Networks (CNNs). Rangesh et. al. \cite{rangesh2016driver} utilizes YOLO \cite{redmon2016you}, a fast ConvNet which enables real-time object detection, and refines detections using a pixel-level skin classifier to segment hand regions. Cronje et. al. \cite{cronje2017training} detects keypoints at the left and right hands using CNN for localizing image regions to detect distraction.

\begin{figure*}
  \centering
  \begin{tabular}{c}
        \includegraphics[width =0.95\textwidth]{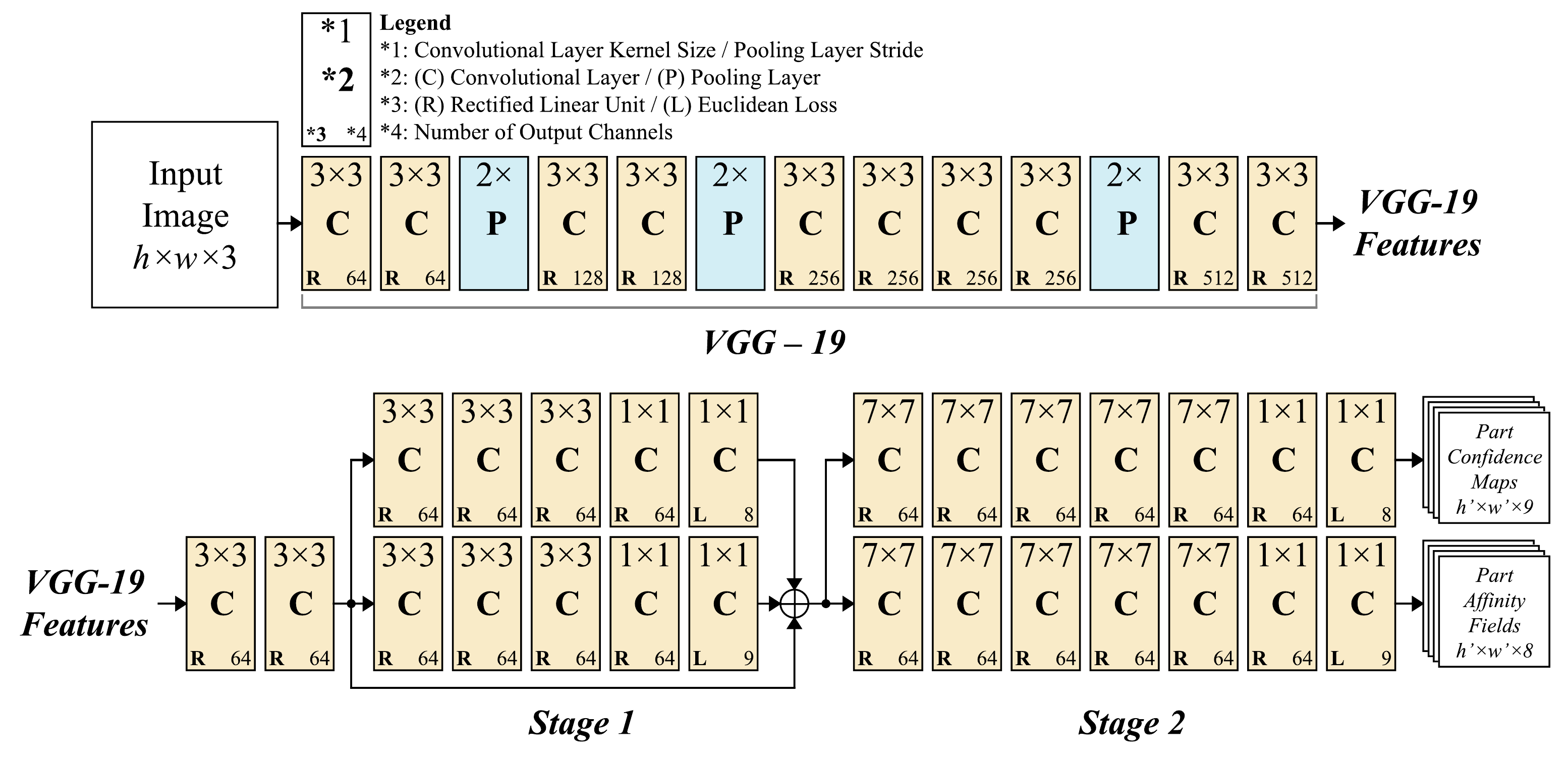}
  \end{tabular}
  \caption{ConvNet architecture, for estimating 8 part confidence maps (top branch), background heatmap where no parts are located (upper branch), and 8 part affinity fields (lower branch), for driver and passenger arms localization. The network takes a 3-channel image as input and passes it through convolutional and pooling layers to output a total of 17 heatmaps at a resolution 8$\times$ smaller than the original input. Euclidean loss is computed with the ground truth heatmaps during training at the end of Stage 1 and 2. Architecture is based on \cite{cao2017realtime}, but utilizies far fewer parameters and only two stages.}
  \vspace{-5mm}
  \label{fig:architecture}
\end{figure*}
There has also been a lot of ongoing efforts and improvement in body pose and joint estimation for people outside of the vehicle. Wei et. al. \cite{wei2016convolutional} designs Convolutional Pose Machine, a sequential multi-stage architecture using convolutional and pooling layers to learn a heatmap for each individual joint, with intermediate supervision at each stage. Newell et. al. \cite{newell2016stacked} performs the same task, designing a multi-stage network using residual modules \cite{he2016deep} along with pooling and upsampling layers, naming it the Stacked Hourglass. Traditionally these heatmaps have been used to learn values from 0 to 1 to indicate the presence of a keypoint. \cite{yuen2017occluded} modified the Stacked Hourglass for the task of learning heatmaps for each of the standard 68 facial keypoints \cite{CMUPIE},\cite{sagonas2016300}. Another modification was also made to the training labels to additional encode occlusion information by outputting values of -1 at the keypoint location if it was occluded and +1 if it was not. Cao, et. al. \cite{cao2017realtime} modifies Convolutional Pose Machines similarly for the Part Affinity Fields in that they range from -1 to 1, encoding angular information for the arm. The Part Affinity Fields provides significant improvement in body joint estimation by using information between each linking pair of joints to improve robustness.

Our research is based on Part Affinity Fields by Cao, et. al., which is a real-time network for body-joint estimation. The primary focus of this work is localizing the hands, and so the network modified to estimate only the wrist-elbow joints. In addition, we significantly reduce the parameters in the later stages providing an even higher framerate of 40 fps, while still performing robustly. The network is trained and evaluated using our own real-world naturalistic data collected in an instrumented testbed equipped with autonomous driving capabilities. We also explore mirror augmentation to reduce annotation time and lighting augmentation to improve robustness to unseen lighting scenarios.

\section{Proposed Approach}


\subsection{Network Architecture}

This work explores and optimizes the network architecture presented in \cite{cao2017realtime} which is selected for the following reasons: 1) heatmap outputs for joints and part affinities which allows, to an extent, a visualization of what the network is learning, 2) real-time performance, 3) well-established network design in the field of body pose and joint estimation with robust results. The network design begins with an input image, which passes through the VGG-19 network utilizing 3 pooling layers with a stride of 2 each, resulting in a large number of feature maps downsampled by 8x resolution. The resulting feature maps are passed through stages of convolutional kernel layers, and the features are also passed directly to the beginning of each stage using a feature concatenation layer. The loss is computed with the ground truth heatmap at the end of each stage, providing an iterative improvement as it progresses through each stage.

The network is designed to simultaenously learn part locations and association between matching parts using Part Affinity Fields (PAFs). The network receives a 3-channel image as input and modified for our application to output 8 part confidence maps, 8 affinity field maps, and a background heatmap. The 8 parts are defined as follows: 1) driver left elbow, 2) driver left wrist, 3) driver right elbow, 4) driver right wrist, 5) front passenger left elbow, 6) front passenger left wrist, 7) front passenger right elbow, 8) front passenger right wrist. Each of the four arms have 2 Part Affinity Field images linking the elbow to the wrist, one contains the x-component of the affinity vector while the other contains the y-component which points along the direction of the arm from the elbow to the wrist. The outputs are visualized in Fig. \ref{fig:intro_fig}.

The reasoning for choosing the elbow to wrist joints is that these two pair of joints form a rigid line due to the human skeleton structure. If elbow to palm-center were to be selected, the line formed between these two joints may not always be along the forearm, which may be more difficult for the affinity maps to learn though we have not experimented on this. An additional joint pair from wrist to palm-center may be a possible option as well, however due to the 8x downsampling, the distance between the two parts may be too small for affinity maps to be taken advantage of fully. We consider the location of the wrist to be close enough to the hand for further analysis, so we follow original body joint locations from the community, namely the wrist and elbow.

The network is modified to utilize only 2 stages and with less than half the parameters in stages 1 \& 2 compared to the original design. The VGG-19 portion of the network remains unmodified so that the network can take advantage of the pre-trained parameters, as done by Cao, et. al. These modifications are made for the following reason: 1) to further speed up processing time where time is a critical factor in the context of driving monitoring, 2) fewer parts to learn (e.g. only elbow-wrist connections), and 3) less variability in image given a fixed camera position in the vehicle cabin. The final architecture with hyperparameters are shown in Fig. \ref{fig:architecture}. 

\subsection{Training Dataset \& Augmentation}
The original network described in \cite{cao2017realtime} was trained using the COCO 2016 keypoints challenge dataset \cite{lin2014microsoft} or MPII human multi-person dataset \cite{andriluka20142d}. As a result, the pre-trained model provided by Cao, et. al. had performance issues running on our data due to the unique camera angle shown in Fig. \ref{fig:intro_fig}. Some of the issues found were: 1) generally no output from OpenPose \cite{cao2017realtime}, 2) confusion of left and right arms in the heatmap outputs, and 3) low scoring output under harsh lighting conditions. In order to resolve these issues, we created our own naturalistic driving dataset with elbow and wrist locations annotated. The images for the dataset were chosen from some of our drives by selecting every 90$^{th}$ frame to cover a wide variety of arm positionings and lighting conditions.

\begin{figure}
  \begin{tabular}{ccc}
    \includegraphics[width=0.95\linewidth]{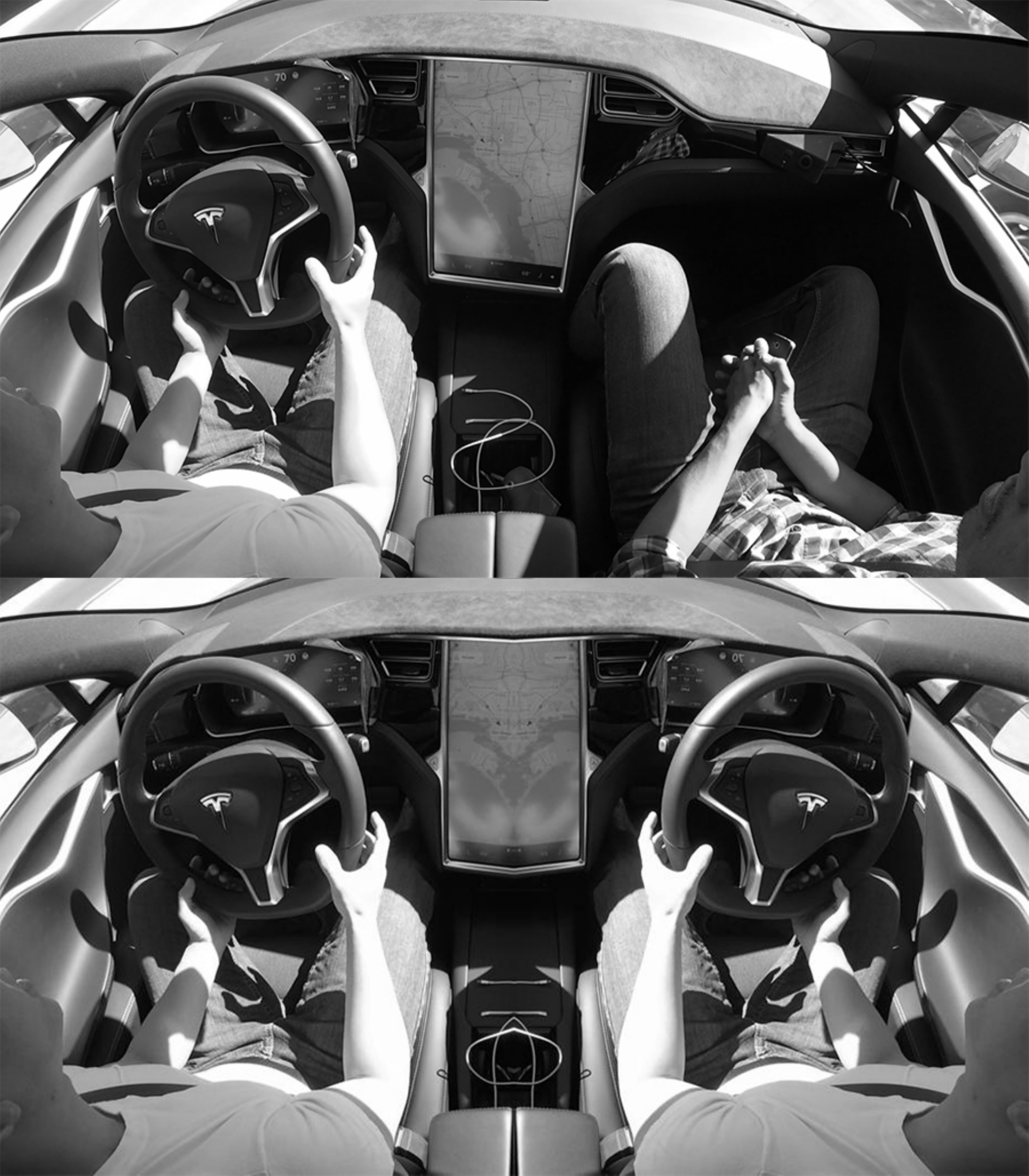}
  \end{tabular}
  \caption{The symmetry-mirror augmentation used to reduce annotation time. (Top) Original camera image with both driver and passenger. (Bottom) Augmented camera image with driver mirrored to passenger side.}
  \vspace{-3mm}
  \label{fig:augmentation_example}
\end{figure}

To reduce the annotation time in half on the training set, only the driver or front passenger arms were annotated to indicate the location of the wrists and elbows parts. During annotation, if the left or right arm were heavily occluded then the image was removed. There are a total of 8500 annotated training images, where approximately 7000 images are annotations of the driver arms only, and the 1500 images are annotations of passenger arms. We take advantage of the vehicle cabin symmetry and mirror the annotated half to the other side which has not been annotated, forming an image of a Driver-Driver or Passenger-Passenger cabin shown in Fig. \ref{fig:augmentation_example}. The annotations are also mirrored to the other side, where driver is specified to be on the left side while the passenger is on the right side for purposes of differentiating between driver and passenger arms with this type of augmentation. In addition, because of limited pool of drivers in the training set with few clothing variations, the images are converted from RGB to grayscale. However, the input images to the network remain as 3-channel grayscale image instead of 1-channel so that the VGG-19 portion of the network does not need to be trained from scratch.

In the vehicle domain, the driver's arms may undergo harsh lighting caused by the sun. It is important that the system remains robust in these situations in order to continue monitoring the driver's hands and behavior or send a signal when it is no longer able to monitor it. To increase robustness to lighting, a lighting augmentation is applied to the images by generating a cloud-like texture image then overlaying it on top of the image. The cloud-like texture is generated using the code provided in \cite{li2013multiple}, then filtered with a Gaussian kernel to blur hard edges away. Other methods of generating cloud-like patterns are also applicable as long as it is different each time it is generated. The texture image is then overlayed with the original image using the following equation from \cite{photoshopformulas}:
\begin{multline}\label{eq:ch4_gauss_eq}
I_i' = \Big(I_i \leq 0.5\Big)\Big(2I_i I_o\Big) + \\ \Big(I_i > 0.5\Big)\Big(1 - 2\big(1-I_i\big)\big(1-I_o \big) \Big)
\end{multline}
where $I_i$ is the input image, $I_o$ is the overlay texture image, and $I_i'$ is the light-augmented image. Here we omit $(x,y)$ for each image variable, e.g. $I_i$ refers to $I_i(x,y)$ indicating the pixel at $(x,y)$.

\begin{figure}
  \begin{tabular}{ccc}
    \includegraphics[width=0.95\linewidth]{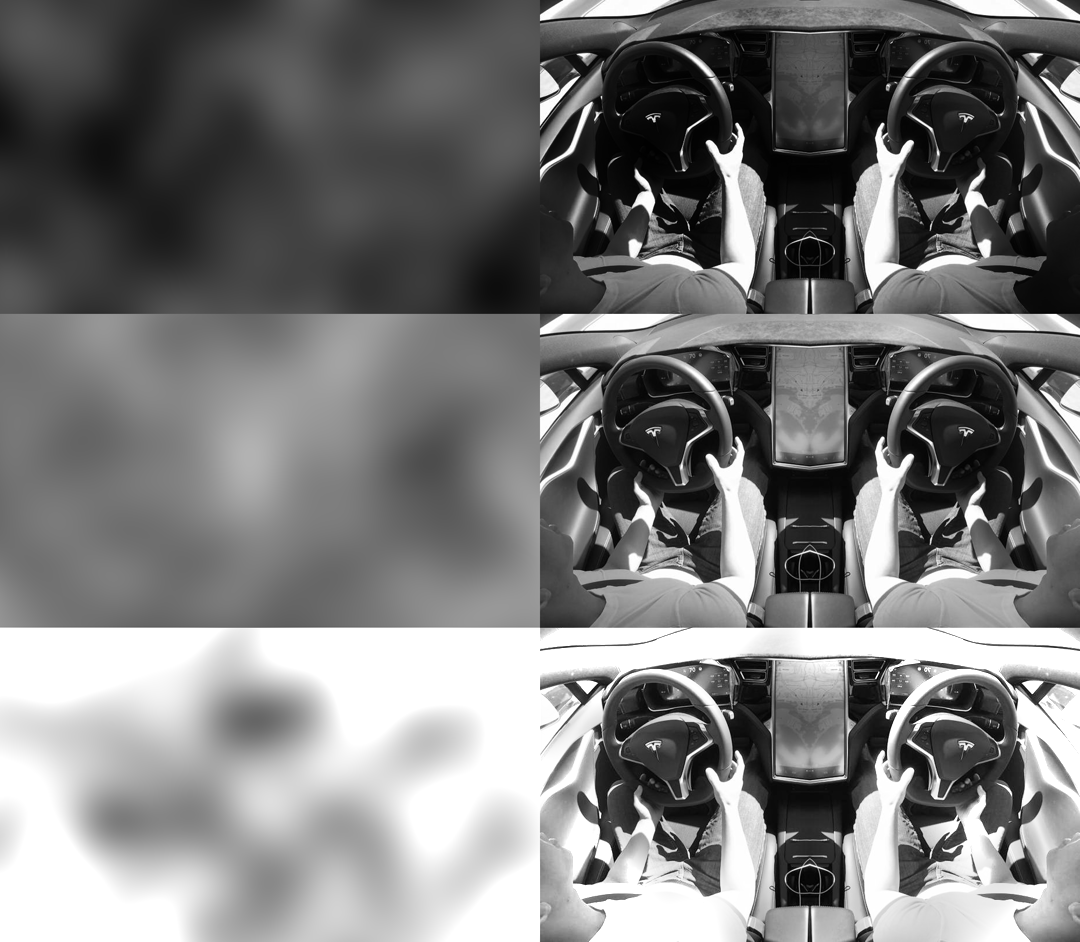}
  \end{tabular}
  \caption{Cloud-overlay augmentation to increase robustness to varying lighting conditions, applied to the same input image. Left and right column shows the artificial cloud images and resulting augmented image, respectively. The top, middle, and bottom rows show the resulting dark, average, and bright augmented lighting conditions, respectively.}
  \vspace{-3mm}
  \label{fig:augmentation_example_lighting}
\end{figure}

In this paper, we focus on three lighting conditions: 1) bright and harsh lighting conditions caused by the sunlight, 2) dark or dim lighting which may occur while driving under freeway underpasses, and 3) regular lighting conditions between bright and dark. To generate these lighting conditions, we scale the pixel intensity values of the cloud texture image such it falls in the range of $[a,b]$. For the three lighting conditions above, we use the following ranges: 1) $[0.4,1.4]$, 2) $[0.05,0.4]$, and 3) $[0.3,0.7]$. Note that an valid image format falls in the range of $[0,1]$, so in the bright lighting condition where the upper limit is $1.4$, any pixels above $1.0$ will be limited to $1.0$. The reason for setting the value at 1.4 is to mimic the saturation effect of the CMOS camera sensors in the situations where the sunlight completely washes out certain areas of the image into blotches of white pixels. An example augmentation is shown in Fig. \ref{fig:augmentation_example_lighting}.

\subsection{Training Label Image Generation}
The training label image generation follows the method described in Cao, et. al., but for completeness we repeat the definitions here. For each of the 8 key points, a PCM image is generated by Eq. \ref{eq:some_gaussian_equation} which essential places a Gaussian blob at the location of the key point, where values in the image range from 0 to 1, highest value indicating the location of the key point.

\begin{equation}\label{eq:some_gaussian_equation}
I_{PCM_i}(\mathbf{x}) = \exp\left(- \frac{||\mathbf{x}-\mathbf{x}_{p_i}||^2_2}{\sigma_{PCM}^2}\right)
\end{equation}
where $\mathbf{x} \in {\rm I\!R}^2$ represents the pixel coordinates in the image and $\mathbf{x}_{p_i} \in {\rm I\!R}^2$ corresponds to the location of keypoint $p_i$.

Two PAF images are generated for each of the four arms using Eq. \ref{eq:some_vector_equation}, one for x-component and other for y-component. The two PAFs represent a vector field along the arm pointing from the elbow to wrist key point. Each pixel in the 2 PAF images represents a vector and their magnitude ranges from 0 to 1, where 1 indicates points along the line formed between the wrist and elbow. Here we define a PAF image as:

\begin{equation}\label{eq:some_vector_equation}
\mathbf{I}_{PAF_j}(\mathbf{x}) = \begin{cases}
    \mathbf{v}, & \text{if $\mathbf{x}$ on arm $j$}.\\
    \mathbf{0}, & \text{otherwise}.
\end{cases}
\end{equation}
\begin{equation}
\mathbf{v} = \frac{\mathbf{x}_{w_j}-\mathbf{x}_{e_j}}{||\mathbf{x}_{w_j}-\mathbf{x}_{e_j}||_2}
\end{equation}
where $\mathbf{x} \in {\rm I\!R}^2$ represents the pixel coordinates in the image and $\mathbf{x}_{w_j},\mathbf{x}_{e_j} \in {\rm I\!R}^2$ corresponds to the location of the wrist and elbow of arm j. Here $\mathbf{x}$ is arm $j$ when the point lies within a distance threshold $\sigma_{PAF}$ from the line formed between the elbow and wrist, defined formally as the set of points $\mathbf{x}$ that satifies the following:\\
$0 \leq \mathbf{v} \cdot (\mathbf{x}-\mathbf{x}_{w_j}) \leq ||\mathbf{x}_{w_j}-\mathbf{x}_{e_j}||_2$ and $|v_\perp \cdot (\mathbf{x}-\mathbf{x}_{w_j})| \leq \sigma_{PAF}$

In the code implementation by Cao, et. al., a background heat map is also learned in the upper branch of the network. This background heat map is essentially the inverse image of the 8 PCMs combined, where the image values range from 0 (keypoint area) to 1 (background area).

\subsection{Training Details}
The network shown in Fig. \ref{fig:architecture} is trained using Caffe \cite{jia2014caffe} and the publicly available code provided by Cao, et. al. \cite{cao2017realtime} with the augmented training images and training label images generated using the 8-key point annotations for both driver and passenger side. The training code also provides a custom datalayer that allows various transformation parameters such as rotation, window cropping, and scaling. For robustness to variation, the images are augmented up to 20 degrees in either direction and scaled from 0.7 to 1.2. Since the network is being trained to differentiate between driver and passenger arms, the crop window width is increased so that the window is large enough to contain all four arms. The network training parameters are listed in Table \ref{tbl:caffe_params}. The network was trained for 160,000 iterations, which took roughly 1-2 days on a GeForce GTX 1080 GPU. 

\begin{table}[]
\centering
\begin{tabular}{|l|l|l|l|l|}
\cline{1-2} \cline{4-5}
Base Learn Rate & 0.00004 &  & Max Rotation     & $\pm$20$^\circ$  \\ \cline{1-2} \cline{4-5} 
Momentum        & 0.9     &  & Crop Width  & 736 \\ \cline{1-2} \cline{4-5} 
Weight Decay    & 0.0005  &  & Crop Height & 368 \\ \cline{1-2} \cline{4-5} 
Gamma           & 0.333   &  & Scale Min        & 0.7 \\ \cline{1-2} \cline{4-5} 
Step Size       & 30000   &  & Scale Max        & 1.2 \\ \cline{1-2} \cline{4-5} 
\end{tabular}
\caption{Network Training Parameters. Parameters used for training the proposed network. (Left) Solver parameters used in Caffe. Parameters used are based on \cite{zhecposegithub}, with the step size adjusted for the proposed dataset used to train the network. (Right) Online augmentation parameters with parameters adjusted with limited rotation and scaling ranges due to the constrained camera position. Crop width is doubled to encompass both driver and passenger.}
\label{tbl:caffe_params}
\end{table}

\begin{figure*}
  \centering
  \begin{tabular}{c}
        \includegraphics[width =0.95\textwidth]{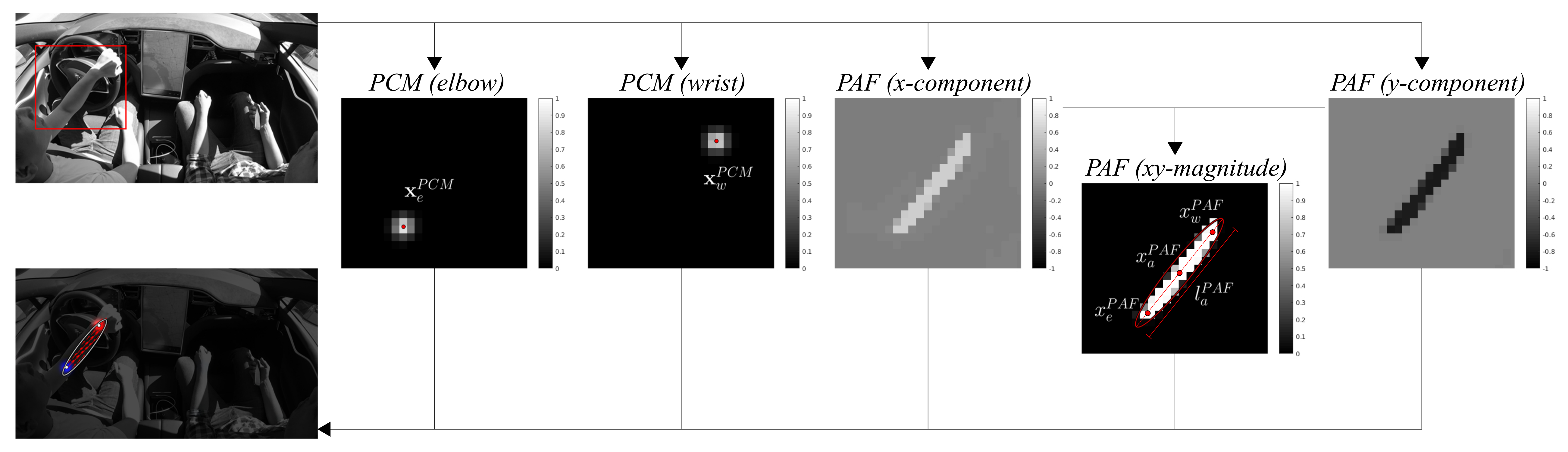}
  \end{tabular}
  \caption{Processing of PCMs and PAFs for the driver's left arm. (Upper left) input image into the network. (Columns 2,3,4,6) PCM \& PAF network outputs for the given input image, cropped to show a close-up in the area of the driver's left arm indicated by the region marked in red in the input image. (Column 5) Magnitude of the x-y PAFs. (Lower left) visualization of the network output for the driver's left arm, overlaying the original input image.}
  \vspace{-5mm}
  \label{fig:hand_processing}
\end{figure*}


\section{Evaluation}
\subsection{Extraction of Part Confidence Maps (PCMs)}
Wrist and elbow coordinates are extracted from PCMs. For a given network output PCM image, we threshold the image by $\tau_{PCM}=0.1$ into a binary image then label connected components. For an ideal network output of the PCM image, there would only be one labeled component. Once the binary image has been labeled, image region properties can be extracted for each labeled region, which is a library tool readily available in MATLAB (e.g. regionprops) and Python (e.g. scikit-image's regionprops). The properties of interest for the PCMs are the weighted centroid and max intensity which represent the coordinate location and the score of the part, respectively. For a given arm (e.g. driver's left arm), we denote the set of possible coordinates and scores of the wrist extracted from the corresponding arm's wrist PCM as $\{\mathbf{x}_{w.PCM}^{1}, ..., \mathbf{x}_{w.PCM}^{n_w}\}$ and $\{S_{w.PCM}^{1}, ..., S_{w.PCM}^{n_w}\}$, respectively. We use the same denotion for elbows, with $e$ in place of $w$, e.g. $\mathbf{x}_{e.PCM}^{n_e}$. Here $n_w$ and $n_e$ are the number of detections or labeled regions found for the wrist and elbow, respectively. An example is shown in Fig. \ref{fig:hand_processing}.

\subsection{Extraction of Part Affinity Fields (PAFs)}
Arm information is extracted from PAFs. The magnitude image of the PAF is computed from the xy PAFs of the corresponding arm. This magnitude image is processed similarly to the PCMs, thresholding the image by $\tau_{PAF}=0.1$, labeling the binary image, then extracting image region properties for each labeled regions. The properties of interest in this case are the ellipse parameters, i.e. centroid, major axis length, and angle. The set of labeled pixels coordinates for a given labeled region are extracted and used to compute the angle and magnitude from the corresponding PAFs. From the set of pixels for the, the median of the angles and magnitude are denoted as the angle of the arm $\theta_{a}^{i}$ and score of the arm $S_{a}^{i}$, respectively, where $i\in\{1,...,n_a\}$ and $n_a$ is the number of detections or labeled regions found for the magnitude PAF of a given arm, e.g. driver's left arm.

In situations where an wrist or elbow part is occluded and scores poorly in the PCM, a secondary estimate of the wrist or elbow can be provided using the estimated ellipse. The PAF estimates of the wrist and elbow locations, based on the ellipse, are given as:
\begin{equation}
	\mathbf{x}_{w.PAF}^{i} = \mathbf{x}_{E}^{i} + \frac{\lambda_aL_a^{i}}{2}\mathbf{n}_{E}^{i}
\end{equation}
\begin{equation}
	\mathbf{x}_{e.PAF}^{i} = \mathbf{x}_{E}^{i} - \frac{\lambda_aL_a^{i}}{2}\mathbf{n}_{E}^{i}
\end{equation}
where $i\in\{1,...,n_a\}$, $\mathbf{x}_{E}^{i}$ is the centroid of the ellipse, $\mathbf{n}_{E}^{i}$ is the unit normal along the ellipse's major axis in the direction of the wrist, $L_a^{i}$ is the length of the ellipse's major axis, and $\lambda_a=0.75$ is a scaling parameter to adjust the estimated location of the wrist from the PAFs along the major axis of the ellipse. The corresponding scores for these PAF-based wrist and elbow ecoordinate estimates are given as $S_{w.PAF}^{i}=\lambda_sS_{a}^{i}$ and $S_{e.PAF}^{i}=\lambda_sS_{a}^{i}$, where $\lambda_s$ is a score adjustment factor. Here we set $\lambda_s=0.5$ to provide the scoring method, described in the next subsection, more incentive to choose the PCM-based wrist/elbow locations over the PAF-based location. An example is shown in Fig. \ref{fig:hand_processing}.

\subsection{Association \& Scoring}
Given a detected arm or ellipse from the PAFs, the wrist is expected to be near a location, $\mathbf{x}_{w.PAF}^i$, based on the geometry of the ellipse. If there is a PCM-based wrist, $\mathbf{x}_{w.PCM}^{j}$, extracted near that location, then it should be used. However, if the PCM-based wrist is too far from the expected location, use the expected location, $\mathbf{x}_{w.PAF}$. From the set of PAF-based and PCM-based wrist estimates, the wrist location is decided using the follow equations:
\begin{equation}
	\mathbf{x}_w^i = \operatorname*{argmax}_{\mathbf{x}} f(\mathbf{x})
\end{equation}
\begin{equation}
f(\mathbf{x}) = \bigg[S_\mathbf{x} \cdot \text{exp}\bigg(-\frac{||\mathbf{x-x_{w.PAF}}||^2}{\sigma_s^2}\bigg) \bigg]
\end{equation}
where $\mathbf{x}\in\{\mathbf{x}_{w.PAF},\mathbf{x}_{w.PCM}^1,...,\mathbf{x}_{w.PCM}^{n_w}\}$, $S_\mathbf{x}$ is the corresponding score of the wrist, and the exponential computes a distance function that decays the function value as the wrist location deviates from the expected wrist location. The updated score for the selected wrist with distance factored in is now defined as:
\begin{equation}
	\mathbf{S_w^i} = f(\mathbf{x}_w^i)
\end{equation}
The elbow is selected using the same formulation described above.

Given an arm-class (e.g. driver's left arm), each arm detection for this particular class is then scored using the following equation:
\begin{equation}
S_{Total}^i = \frac{1}{3}\left(S_a^i + S_w^i + S_e^i\right)
\end{equation}
where $S_a^i$, $S_w^i$, and $S_e^i$ are the individual component scores of the arm, wrist, and elbow components. The highest scoring detected arm is then chosen as the driver's left arm, providing the highest scoring estimate for the wrist-elbow part locations. The processing of PCMs, PAFs, and association/scoring is then repeated for the remaining 3 arms (driver's right arm, passenger's left arm, and passenger's right arm).

The framework presented in this paper takes a different approach to Cao, et. al.'s work on the usage and processing of the affinity fields. In our experiments, we found that it is more likely for the joint heatmaps to have lower output scores due to self-occlusion from the head/shoulders or from objects being held. As a result, there are situations where the joint's location may not be determined from the joint heatmap, however the affininity fields showed high score results in such situations allowing for a rough location of the joint to be estimated from the PAFs if the PCMs failed. This is because the arm is more difficult to occlude the entire arm compared to occluding a smaller area such as the elbow or wrist. In this work, the localization of the arm begins with the affininity fields instead of the joint maps, whereas the localization of the body joints begins with joint maps in the original work.

\begin{figure*}
 \centering
 \begin{tabular}{c}
       \includegraphics[width =1\textwidth]{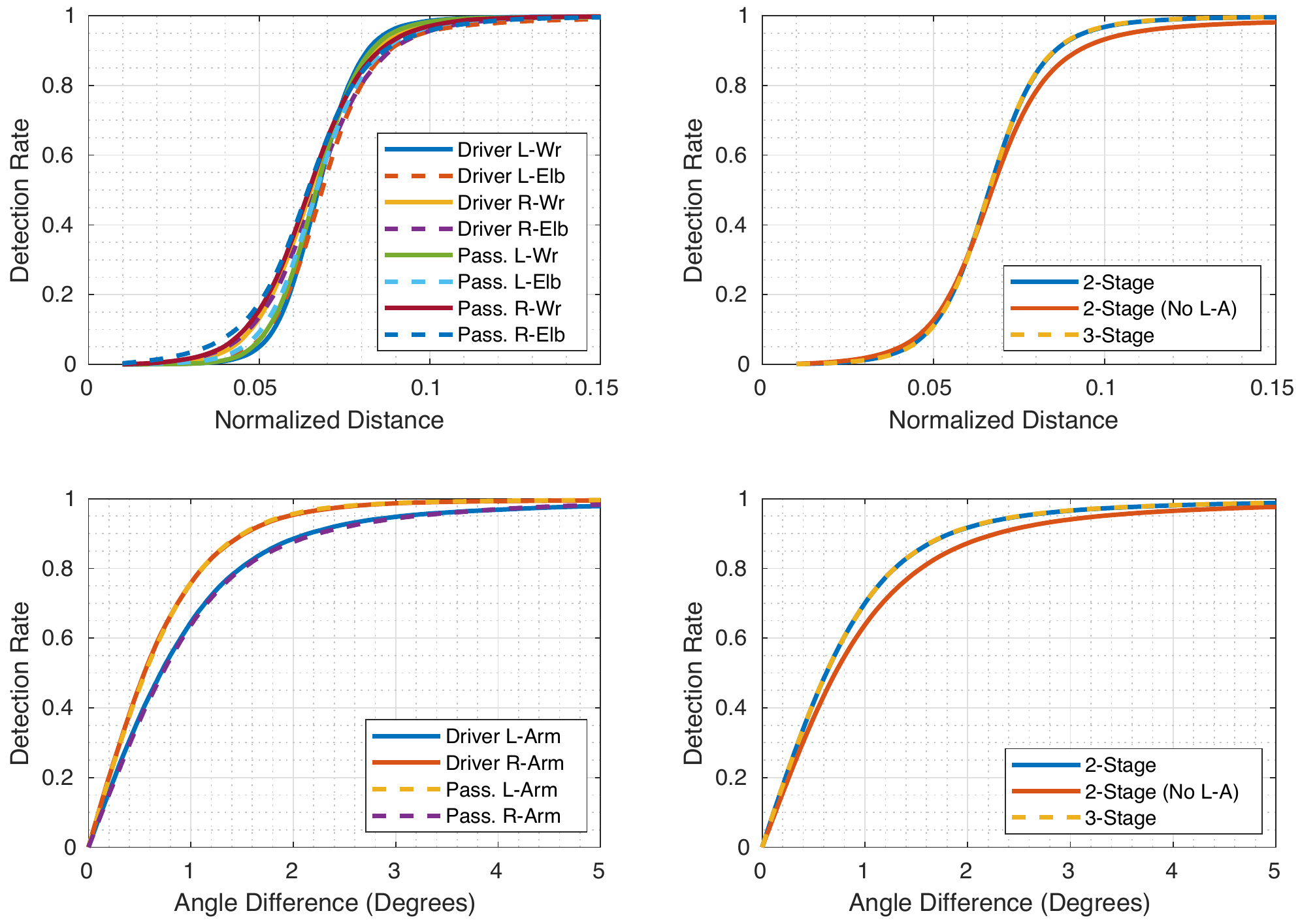}
 \end{tabular}
	\caption{Evaluation on our naturalistic driving test dataset. (Top left) PCK curve showing 2-stage with lighting augmentation (L-A) performance of the 8 parts, normalized by average length segment of all arms in the dataset. (Top right) PCK curve of all parts comparing performance between 2-stage with \& without L-A, and 3-stage with L-A. (Lower left) Angle performance curve of 2-stage with L-A showing the detection rate median angle estimation of the Part Affinity Fields for each arm, where detection rate is the percentage of images where the estimated angle is below the specified absolute angle difference threshold. (Lower right) Angle performance curve for all arms comparing between between number of stages and augmentation.}
 \vspace{-5mm}
 \label{fig:result_curves}
\end{figure*}

\subsection{Test Dataset \& Results on Real World Autonomous Drive Data}
The test set is composed of 1500 images sampled from multiple drives in our naturalistic day-time driving dataset which includes some drivers found in the training set and also new unseen drivers. For evaluation, the images are annotated with the location of the wrist and elbow joints for both driver and passenger arms, unlike the training set which has only the driver or the passenger annotated. The test set is evaluated using the Percentage of Correct Keypoints (PCK) metric, which provides the percentage of detections that meet the normalized distance threshold of the ground truth. This metric is a standard in the Body Pose and Joint estimation community, where it is generally normalized by the size of the torso or the head segment. We change our method of normalization for two reasons: (1) the camera is at a fixed distance and location, so we can use a constant normalization value across all images, (2) since there are no torso/head segments annotated, we normalize using the average length of all arms in our ground truth dataset. The results for the 8 part's localization performance are shown in Fig. \ref{fig:result_curves}a, reporting 95\% detection rate for each part at a distance threshold of 10\% of the average arm length, measured from wrist to elbow, in the dataset.

Using the same idea behind the PCK metric, the median angle computed from the PAF maps for each arm is also evaluated. The ground truth angle of the arm is computed by:
\begin{equation}
\theta_{GT}=tan^{-1}\left(\frac{y_w-y_e}{x_w-y_e}\right)
\end{equation}
where $(x_w,y_w)$ and $(x_e,y_e)$ are the annotated coordinates of the wrist and elbow, respectively. The metric represents the percentage of detections that fall within the absolute angle difference threshold. The results for estimating the angle for each of the four arms are shown in Fig. \ref{fig:result_curves}c. This shows that at least 95\% of the results are within 5 degrees of the ground truth angle.

PCK and angle performance curves comparative analysis are shown Fig. \ref{fig:result_curves}b,d, where all 8 joints and four arms are analyzed together instead of separately. Comparison is shown for training the 2-stage network with and without lighting augmentation (``No L-A'' in legend), showing a noticable increase in performance for both joint localization and angle estimation. The graphs also show the comparison between 2-stage and 3-stage networks trained with lighting augmentation, resulting in little to no performance gain from 2-stage to 3-stage for both joint localization and angle estimation. As a result of these experiments, our application will utilize the 2-stage network trained with lighting augmentation so that it can run at a higher framerate speed and/or utilize less GPU resources.

\begin{figure*}
 \centering
 \begin{tabular}{c}
       \includegraphics[width =0.95\textwidth]{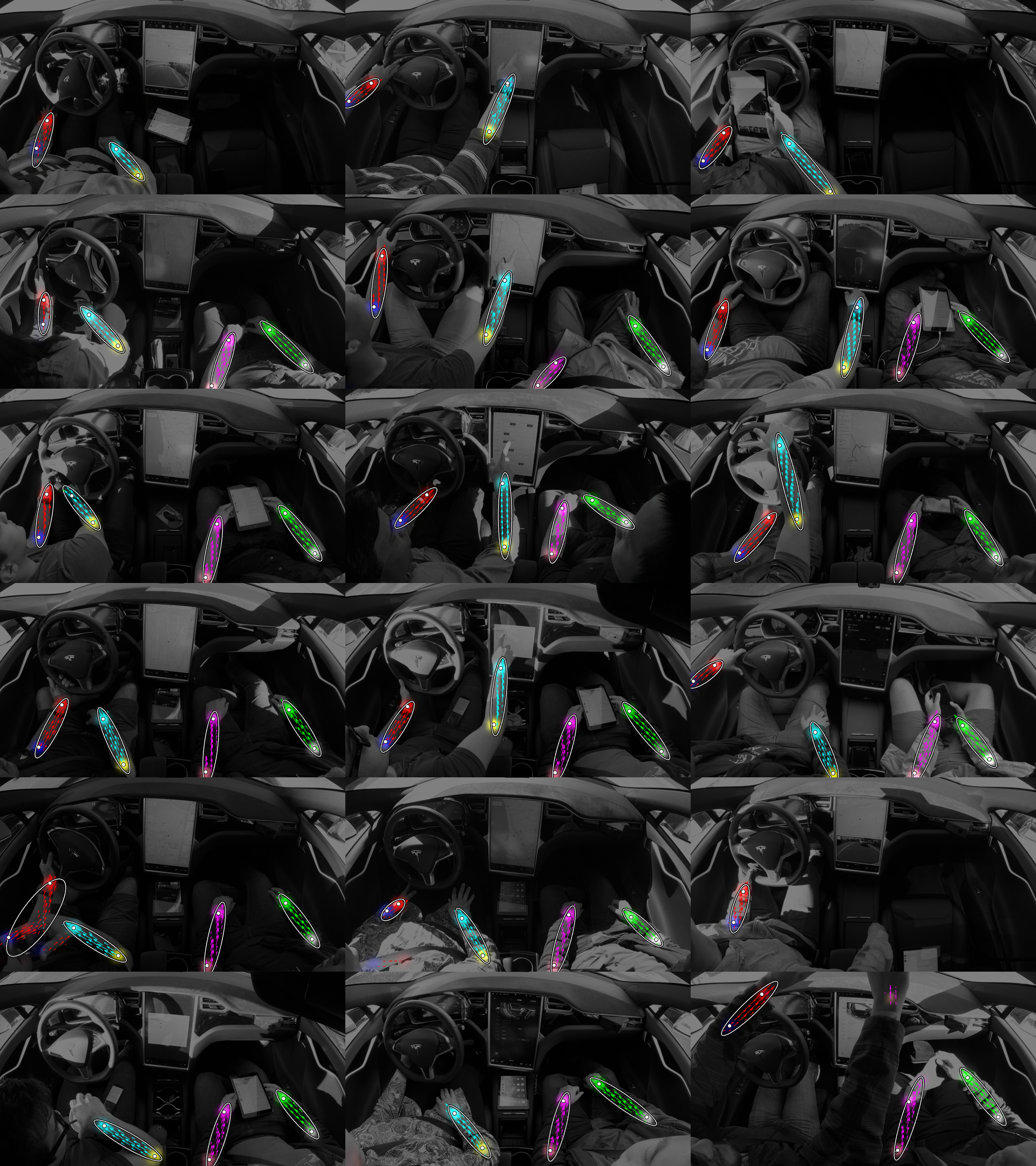}
 \end{tabular}
	\caption{Visualizing the results of Part Confidence Maps and Part Affinity Fields generated by the network. Each color is a different channel output, corresponding to each of the driver and passenger arms. The white points represent the estimated part locations. Ellipses represent the estimated ellipse over the vector field. (First Row): Without any front-seat passengers, showing very low heatmaps for passenger area side when it's empty. (Last Two Rows): Failure cases where it does not perform as well due to unseen occlusions or unique clothing.}
 \vspace{-5mm}
 \label{fig:result_images}
\end{figure*}

Example results from our network are shown in Fig. \ref{fig:result_images}. Different drivers and passengers are shown, with various arm positioning and gestures. Examples are shown in situations where there is no front passenger, showing that the network outputs very low scores when there is no passenger. The bottom two rows shows some failure cases. The first column of the last two rows shows issues with self-occlusion, where the driver's right arm or head blocks the view of the left arm. The upper image shows that the driver's right hand is blocking the driver's left elbow, causing issues with the network's output heatmaps driver's left joint heatmaps. The second column of the last two rows shows a driver with unique clothing with heavy artistic texturing that has not been seen in the training set before, resulting in poor performance for the left arm. The right column shows some failure cases where the driver is reaching for an object in the cupholder area or adjusting rearview mirror. As a result of the mirror-augmentation, arms closer to the center may not be detected well. Some of the issues seen here can be resolved by: 1) placing the camera in a more strategic location to avoid head occlusions or using multiple camera views, 2) increase training dataset with more variety in clothing, 3) add training samples without the use of mirror-augmentation so that the network sees examples of arms directly along or past the center of the image, where the mirror occurs.

\begin{figure*}
 \centering
 \begin{tabular}{c}
       \includegraphics[width =0.95\textwidth]{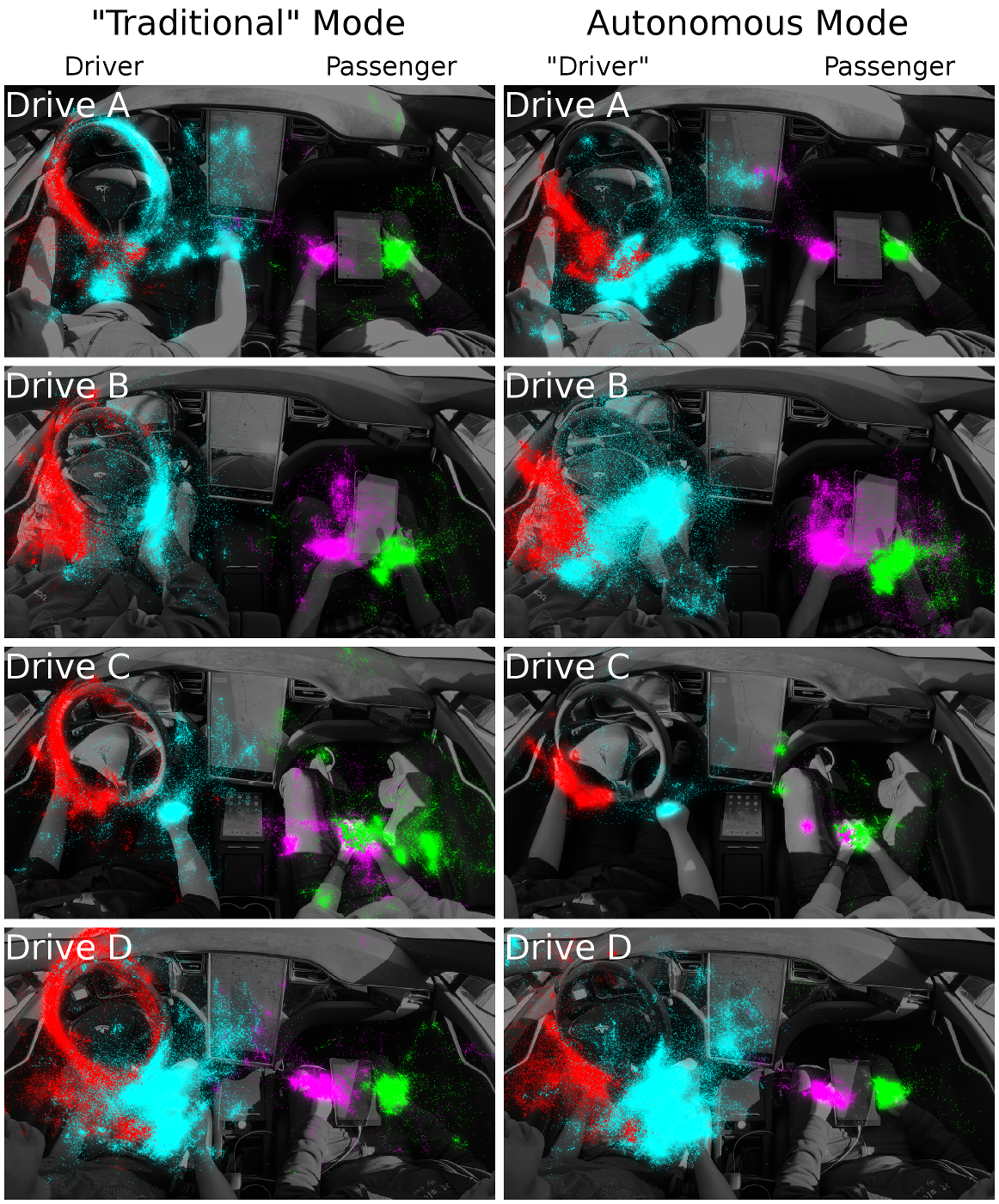}
 \end{tabular}
	\caption{Visualization of the location of the hands, extrapolated using the estimated locations of wrist and elbow, over long drives ranging from 30-120 minutes in duration in our naturalistic driving dataset collected in an autonomous-capable vehicle. Each color represents one of the four hands in the camera view. Plotted points represent all instances of the hand location throughout the drive, while the colored heatmap or glow represents how often the hand has been in the given area. Four different drives are shown, one drive per row. (Left column): Instances where the vehicle is in traditional (or manual) driving mode or Traffic Aware Cruise Control (TACC) mode, where steering is controlled by driver hands. (Right column): Instances where the vehicle is in autonomous mode with auto-steering and TACC enabled, where the driver does not need to control foot pedals or steering wheel but is required to pay attention.}
 \vspace{-5mm}
 \label{fig:manual_vs_auto}
\end{figure*}

An approximate hand location can be computed similar to \cite{rangesh2016pedestrians} using the following equation:
\begin{equation}
\mathbf{x_h}=\mathbf{x_w}+\lambda_h\left(\mathbf{x_w}-\mathbf{x_e}\right)
\end{equation}
where $\mathbf{x_h}$, $\mathbf{x_w}$, $\mathbf{x_e}$ are the locations of the hand, wrist, and elbow, respectively. The second term represents the a vector in the direction from the elbow to wrist, where the length of the vector is represents the length of the arm scaled by a factor $\lambda_h=0.25$ which generates the best results in this work. By approximating the hand location in this way, the driver and passenger hand locations can be analyzed over time as shown in Fig. \ref{fig:manual_vs_auto}, similarly done in \cite{ohn2014beyond}. The figure shows analysis for four different drives with varying drivers and passengers, where the vehicle is capable of autonomously driving itself. In current vehicles, the passengers generally has little to no control compared to the driver as shown by the analysis where the passenger hands generally remain stationary in a general location. In autonomous mode where the vehicle controls the steering of the vehicle, the driver's left hand is seen more concentrated on the lower left of the steering wheel. One reason for this is due to the vehicle requesting the driver to hold the steering wheel at irregular intervals through a message on the speedometer, where the system will eventually disable autonomous mode in a safe manner if the driver still has not touched the steering wheel after multiple messages. In manual mode, the left hand may cover a lot more area on the steering wheel due to making turns since the vehicle currently cannot make left or right turns in autonomous mode. In some drives, the driver is seen to move their right hand in various locations a lot more while in autonomous mode compared to manual mode as seen in the first two rows of Fig. \ref{fig:manual_vs_auto}, performing secondary activites or hand gestures while talking to the passenger. Depending on the driver and the drive itself, the driver may not change the movement habits of their right hand by much as shown in the last two rows between the two different vehicle modes. 



\section{Concluding Remarks}
In this paper, a framework has been proposed to localize and classify four different arms with a total of 8 joints based on the work by Cao, et. al. The original ConvNet is retrained with a significant parameter reduction and using training data with a unique camera angle on the human body for in-vehicle occupant monitoring. Two augmentation methods are presented in this paper. The first method is mirroring the annotated half of the image (e.g. driver side) to the un-annotated side (e.g. passenger side) allows us to cut annotation time in half for training images, or double the amount of training data if both sides were annotated by applying the mirroring for each side. The second method is the augmentation of lighting to improve the network's robustness to variation in lighting by generating randomized cloud texture images varying from dark to light and overlaying it on top of the training image.

Our method shows optimistic performance with 95\% detection rate for the localization on each of the 8 joints, with a localization error of up to 10\% of the average arm length, and also 95\% detection rate in estimating the angle of the arm from the PAFs to within 5 degrees. The failure cases shown in the evaluation section may be resolved by retraining the network with a lot more variation and clothing. This can be done in a semi-supervised manner by running the network on videos and automatically selecting images with low-scoring results, which will then be annotated for the network to re-learn and provide higher-scoring results in the future, allowing the network to iteratively improve in performance as more data is collected. The arm detector in this framework will be used as the foundation for future work in monitoring driver activity such as analyzing the hand behaviors in manual driving mode compared to autonomous mode, hand gesture analysis, and secondary tasks and/or objects.

\section{Acknowledgments}
The authors would like to specially thank Akshay Rangesh and reviewers for their helpful suggestions towards improving this work. The authors would like to thank our sponsors and our colleagues at the Laboratory for Intelligent \& Safe Vehicles (LISA) for their assistance.


%





\ifCLASSOPTIONcaptionsoff
  \newpage
\fi



%


\bibliographystyle{IEEEtran}
\bibliography{bare_jrnl_bib}

%

\begin{IEEEbiography}[{\includegraphics[width=1in,height=1.25in,clip,keepaspectratio]{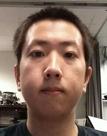}}]{Kevan Yuen} received the B.S. and M.S. degrees in electrical and computer engineering from the University of California, San Diego, La Jolla. During his graduate studies, he was with the Computer Vision and Robotics Research Laboratory, University of California, San Diego. He is currently pursuing a PhD in the field of advanced driver assistance systems with deep learning, in the Laboratory of Intelligent and Safe Automobiles at UCSD.
\end{IEEEbiography}

\begin{IEEEbiography}[{\includegraphics[width=1in,height=1.25in,clip,keepaspectratio]{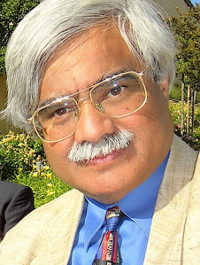}}]{Mohan Manubhai Trivedi} is a Distinguished Professor of and the founding director of the UCSD LISA: Laboratory for Intelligent and Safe Automobiles, winner of the IEEE ITSS Lead Institution Award (2015). Currently, Trivedi and his team are pursuing research in distributed video arrays, active vision, human body modeling and activity analysis, intelligent driver assistance and active safety systems for automobiles. Trivedi’s team has played key roles in several major research initiatives. These include developing an autonomous robotic team for Shinkansen tracks, a human-centered vehicle collision avoidance system, vision based passenger protection system for “smart” airbag deployment and lane/turn/merge intent prediction modules for advanced driver assistance. Some of the professional awards received by him include the IEEE ITS Society’s highest honor “Outstanding Research Award” in 2013, Pioneer Award (Technical Activities) and Meritorious Service Award by the IEEE Computer Society, and Distinguished Alumni Award by the Utah State University. Three of his students were awarded “Best Dissertation Awards” by professional societies and a number of his papers have won “Best” or “Honorable Mention” awards at international conferences. Trivedi is a Fellow of the IEEE, IAPR and SPIE. Trivedi regularly serves as a consultant to industry and government agencies in the U.S., Europe, and Asia.
\end{IEEEbiography}

\end{document}